\makeatletter
\def\vereq#1#2{\lower3pt\vbox{\baselineskip1.5pt \lineskip1.5pt
\ialign{$\m@th#1\hfill##\hfil$\crcr#2\crcr\sim\crcr}}}

\makeatother

\newcommand{\bm}[1]{\mbox{\boldmath$#1$}}

\documentclass{jpsj2}
\title{Ground-State Properties of a Heisenberg Spin Glass Model 
with a Hybrid Genetic Algorithm}

\author{Yuh-ichi Iyama\dag and Fumitaka Matsubara}

\inst{%
Department of Applied Physics, Tohoku University, Sendai 980-8579,
Japan}

\recdate{\today}

\abst{%
We developed a genetic algorithm (GA) in the Heisenberg model 
that combines a triadic crossover and a parameter-free genetic algorithm. 
Using the algorithm, we examined the ground-state stiffness of 
the $\pm J$ Heisenberg model in three dimensions up to a moderate size range. 
Results showed the stiffness constant of $\theta = 0$ 
in the periodic-antiperiodic boundary condition method 
and that of $\theta \sim 0.62$ in the open-boundary-twist method. 
We considered the origin of the difference in $\theta$ between the 
two methods and suggested that both results show the same thing: 
the ground state of the open system is stable against a weak perturbation. 
}

\kword{%
genetic algorithm, ground state, spin-glass}

\begin{document}
\maketitle

\section{Introduction}

Optimization methods have found widespread application 
in computational physics. Among these, the investigation of the 
low-temperature behavior of spin glasses (SGs) has attracted much attention 
within the statistical physics community
because, despite its simple definition, its behavior is far from understood. 
Particularly, domain-wall energies of the SGs at the absolute zero temperature 
($T=0$) are suggested to give a stable SG phase at a very low temperature. 
From a computational point of view, the calculation of spin-glass ground 
states is very demanding and various algorithms have been 
developed\cite{Algorithms}. 
P\'{a}l found that genetic algorithms (GAs) are useful in searching for 
the ground state of the $\pm J$ Ising model\cite{Pal}. He proposed a triadic 
crossover together with an effective local optimization method. 
The GA algorithm has 
been improved in several ways; a powerful local optimization method is 
attached\cite{Hartmann}, and the GA is coded in a different 
manner\cite{Houdayer1,Houdayer2}. The system size that is treatable in a 3D 
SG model is moderate at $13 \times 13 \times 13$. 
Unfortunately, those algorithms more or less use the nature of the Ising 
spin and do not apply continuous spin models. 
One of the authors and his coworkers showed that a GA also works 
in the Heisenberg SG model\cite{Takahashi}. They applied a one-point crossover 
together with interface optimization. Although it enables us to treat 
the 3D $\pm J$ model up to a moderate size, its coding was rather complicated; 
several improvements are necessary for application to different systems.

In this paper, we will develop a GA in a different manner for the Heisenberg 
model and consider the ground-state properties of the 3D $\pm J$ Heisenberg 
model on $L \times L \times L$ described by the Hamiltonian 
\begin{equation} 
     H = - \sum_{\langle ij \rangle} J_{ij}\bm{S}_{i}\cdot\bm{S}_{j} , 
\end{equation} 
where $\bm{S}_{i}$ is the Heisenberg spin of $|\bm{S}_i| = 1$ and 
$\langle ij \rangle$ runs over all nearest-neighbor pairs. 
The exchange interaction $J_{ij}$ takes on either $+J$ or $-J$ with 
the same probability of 1/2. 
Using the algorithm, we examine the ground-state stiffness 
of the model using a periodic-antiperiodic boundary method and 
an open-boundary-twist method.

In section 2, we will give two types of GA for the Heisenberg model. 
First, we will show that the triadic crossover in the Ising model 
also works well in the Heisenberg model (GAI). 
Then, combining it with a parameter-free genetic algorithm in 
optimization problems, we will give a useful algorithm (GAII). 
In section 3, using the GAII, the domain-wall energy $\Delta E_L$ of the 
model will be calculated up to a moderate size of the lattice ($L \leq 12$) 
and the stiffness constant $\theta$ will be given in good accuracy 
in both the periodic-antiperiodic and open-boundary-twist methods. 
Section 4 will be devoted to conclusions.

\section{Genetic Algorithm}

We develop two GAs in the Heisenberg model: one is the GA proposed by P\'{a}l 
in the Ising model\cite{Pal}; the other is that proposed in other optimization 
problems.

\subsection{A Triadic Crossover {\rm (GAI)}}

We first consider P\'{a}l's GA, which can be explained briefly as follows. 
One starts with a population of $M_i = 2^{\lambda}$ random configurations 
({\it individuals}) of the Ising spins $\{\sigma_i^{(k)} = \pm1 \}$ 
($k=1, 2, 3, \cdots, M_i$), which are linearly arranged in a ring. Then 
two neighbors from the population 
are taken ({\it parents P1} and {\it P2}) and two offspring are created 
using a triadic crossover: For each offspring, (i) a different 
individual ({\it mask} or {\it reference R}) is selected from the population, 
and the spin of the offspring $\{\sigma_i^{(O)}\}$ is put as $\sigma_i^{(O)} 
= \sigma_i^{(P2)}$ or $\sigma_i^{(O)} = \sigma_i^{(P1)}$ if 
($\sigma_i^{(R)}\sigma_i^{(P1)}) = 1$ or -1, respectively;
(ii) some fractions $\mu$ of the spins of the offspring are reversed 
({\it mutation}); 
(iii) the offspring is optimized using some local optimization method. 
Each offspring competes with only one parent: the one that is more similar 
to the offspring. The parent is replaced by the offspring if the energy of 
the parent is not lower than that of the offspring.
After this step is repeated $\nu_0 M_i$ times ({\it $\nu_0$ is called a 
generation period}), the population is halved to save computer CPU time. 
From each pair of neighbors, the individual with a higher energy is 
eliminated. 
This procedure progresses until only four individuals remain in the last 
stage. One merit of this algorithm is that, using the triadic crossover, 
one can mix the spin configurations of P1 and P2 over the lattice. 
Another merit is its ability to slow the loss of population diversity. 
Both are indispensable for the efficient search for the ground state.

The algorithm might be applied to the Heisenberg model because its components, 
which are characteristic of the Ising model, are the crossover rule and the 
local optimization method. 
We might replace the equality of the crossover rule with an inequality and 
the local optimization method with one appropriate for the Heisenberg model. 
We start with the population of individuals of the Heisenberg spins 
$\{\bm{S}_i^{(k)}\}$. We generalize the crossover rule as
\begin{equation}
\bm{S}^{(O)}_i = \left\{ 
\begin{array}{c} \bm{S}_i^{(P2)} \hspace{0.5cm} {\rm for}\hspace{0.5cm} 
          r_i^{(P1)} >   0 \hspace{0.5cm} , \\
                 \bm{S}_i^{(P1)} \hspace{0.5cm} {\rm for}\hspace{0.5cm} 
          r_i^{(P1)} <   0 \hspace{0.5cm} , 
\end{array} 
\right.
\end{equation}
where $r_i^{(k)} = (\bm{S}_i^{(k)}\bm{S}_i^{\rm (R)})$, $k =$ 
P1 or P2\cite{Comm_Traidic}. 
We use a spin quench method for local optimization; all spins are aligned 
successively in the directions of their local fields. This procedure 
is repeated many times ({\it spin quench step $N_q$}). 
The similarity of two individuals, $k$ and $l$, is measured with a distance 
$\Delta S_{k,l}$ between them: 
\begin{equation} 
\Delta S_{k,l}=\sqrt{\frac{1}{N^2}\sum_{i,j}(\bm{S}_{i}^{(k)}\bm{S}_{j}^{(k)}
                                    - \bm{S}_{i}^{(l)}\bm{S}_{j}^{(l)})^2}.
\end{equation}
Actually, $\Delta S_{k,l} =0$ when the two individuals are equivalent apart 
from a uniform rotation, although $\Delta S_{k,l} \sim \sqrt{2/3}\;
(\equiv \Delta S_{\infty})$\cite{Takahashi} when they are independent. 
Consequently, all components of the P\'{a}l algorithm have been 
prepared in the Heisenberg model and the algorithm is applicable to it. 
We designated this GA as GAI. 
The CPU time $t_{\rm I}$ of the GAI is estimated as
\begin{equation} 
t_{\rm I}=A_{\rm I} \times\nu_0 \times 2(2M_i-1) \times N_q \times({\rm lattice\ size}\ N), 
\end{equation}
where $A_{\rm I}$ is a constant that is independent of other parameters. 
Therefore, the CPU time depends on three parameters, $M_i$, $\nu_0$, 
and $N_q$, in addition to the lattice size $N$. 
In other words, we must determine the values of those parameters as well as 
that of the mutation fraction $\mu$ to optimize the algorithm. 
Fixing $\mu = 0.1$, we test the GAI in the $\pm J$ Heisenberg model on 
the $L \times L \times L$ lattice with periodic boundary conditions. 
For a given bond configuration (sample), we apply the GAI 10 times 
$(n=1,2,3,\cdots,10)$ using different initial populations and obtain the 
lowest energy $E_n$ for each time. 
We tentatively assume the ground-state energy as $E_G = \min\{E_n\}$ and 
presume that we can succeed in getting the ground-state energy when 
$(E_n - E_G)/|E_G|<10^{-6}$. 
We made such a test for $N_s = 16$ samples; results for $L \leq 12$ are 
presented in Table I. 
We see that search efficiency is improved considerably compared with 
that of the conventional spin quench (SQ) method.\cite{Takahashi} 
However, to maintain efficiency, 
we need twice or more of the CPU time when $L \rightarrow L+1$. 
One usually encounters this difficulty in the optimization problem of 
complex systems. 
Here it will be enhanced by eliminating individuals at every 
$``\nu_0"$, which degrades the diversity of the population. 
It is apparent that a large period $\nu_0$ slows the loss. 
In fact, as presented in Table I, search efficiency is improved when 
$\nu_0 \rightarrow 2\nu_0$, but twice the CPU time is necessary.

In summary, the triadic crossover of the Ising model is 
applicable to the Heisenberg model; furthermore, it improves the 
ground-state search efficiency considerably. 
We should optimize many parameters for application to different systems.

\begin{table*}
\label{t1}
\caption{\label{tabone} Ratios and their standard variances of success in
searching for the ground-state energy of the $\pm J$ model on 
$L \times L \times L$ lattice using the GAI, 
where $M_i$, $\nu_0$, and $N_q$ respectively represent the populations in the 
initial stage, the generation period, and the quenching steps. 
These are estimated for 10 runs for 16 samples. } 
\vspace{0.3cm}
\begin{center}
\begin{tabular}{@{\hspace{\tabcolsep}\extracolsep{\fill}}r|cccccc} \hline
$ L (\nu_0,N_q) \setminus  M_i$ & 32 & 64 & 128 & 256 & 512 & 1024 \\ \hline
8  (16,200)   &0.34(.23)&0.52(.23)&0.74(.23)&0.90(.20)&---&--- \\ 
10 (16,200)   &---&0.59(.31)&0.75(.30)&0.87(.19)&---&---  \\
   (32,200)   &---&0.70(.31)&0.86(.24)&0.99(.03)&---&---  \\
11 (32,200)   &---&---&0.51(.20)&0.81(.25)&0.92(.12)&---  \\
   (64,200)   &---&---&0.67(.21)&0.89(.14)&0.98(.04)&---  \\
   (64,400)   &---&---&0.87(.15)&0.97(.04)&0.99(.03)&---  \\
12 (64,400)   &---&---&---&0.81(.23)&0.95(.09)&0.98(.04)  \\
\hline
\end{tabular}
\end{center}
\end{table*}

\subsection{Parameter-free genetic algorithm {\rm (GAII)}}

Next, we consider the {\it ``Parameter-free genetic algorithm''} (PfGA) of 
Sawai and coworkers.\cite{Sawai1,Sawai2,Sawai3}. 
The motivation of the PfGA is to reduce the number of parameters to be 
determined beforehand to apply them to different problems. 
Although the PfGA reduces such parameters, it proves its efficiency 
in a benchmark test on ICEO\cite{ICEO}. 
A characteristic point of the PfGA is to consider local populations. 
The numbers of the individuals in those local populations 
are not the same and vary as the population evolves. 
Another point is the asymmetry of the mutation between two offspring, which 
is inspired by the disparity theory of evolution by Furukawa et 
al.\cite{Furusawa1,Furusawa2}; for one offspring, the mutation fraction 
$\mu$ is given as a random number, although no mutation is applied for the 
other offspring. The PfGA might also be applied the Heisenberg spin system.

We consider the population $S$ of $N_p$ local populations S$^{\prime}_k$ 
($k = 1,2,\cdots,N_p$). The numbers of the individuals in those local 
populations are $n_k$ and the total number of the individuals in the 
population is $n (=\sum_kn_k)$.
As a first step, each local population has two individuals $n_k = 2$ that 
are constructed randomly and optimized. They evolve as follows:
\begin{enumerate}
\item Select a local population S$^{\prime}_k$ randomly according 
to the ratio of $n_k/n$. 

\item Select two individuals randomly from S$^{\prime}_k$ and take out. 
They consist of a family of S$^{\prime}_k$. That is, they are considered 
as the parents P1 and P2 with their energies $E_{\rm P1}$ and 
$E_{\rm P2} (\geq E_{\rm P1})$. 

\item Two offspring (children) are produced in the family as follows. 
\begin{enumerate}
\item Select one individual, R1, from different S$^{\prime}_l$($l \neq k$); 
generate a child, C1, applying the triadic crossover rule described in 
Sec. 2.1. 
Generate another child, C2, applying the same rule with a different 
individual, R2, from another S$^{\prime}_{l^{\prime}}$($l^{\prime} \neq k$).

\item A mutation algorithm is applied to one child (C2): Choose 
a random number, $x_m$, between (0,1).  Attach a different random number, 
$x_i$, to each spin $\bm{S}_i^{(C2)}$.  If $x_i < x_m$, $\bm{S}_i^{(C2)}$ 
is replaced by a new one, which is constructed randomly; otherwise, 
it remains unchanged.

\item The spin quench algorithm is applied for $N_q$ 
steps to optimize C1 and C2. The energies of C1 and C2 are $E_{\rm C1}$ and 
$E_{\rm C2}$. If $E_{\rm C1} > E_{\rm C2}$, then the spin configurations of 
C1 and C2 are mutually exchanged. 
\end{enumerate}

\item Select $1 - 3$ members from the family and return to S$^{\prime}_k$. 
The selection rule is as follows. 
\begin{enumerate}
\item If $E_{\rm C2} \leq E_{\rm P1}$, we choose C1, C2 and P1. 
Here, we consider that P1 has superior genes and it still has the ability 
of yielding another good child. 
\item If $E_{\rm P2} \leq E_{\rm C1}$, we choose P1. 
Here, we consider that no superior gene exists in P1 or in P2. 
\item If $E_{\rm C1} \leq E_{\rm P1} \leq (E_{\rm C2} \leq E_{\rm P2}$ 
{\rm or} $\leq E_{\rm P2} < E_{\rm C2}$), we choose C1 and P1. 
\item If $E_{\rm P1} \leq E_{\rm C1} \leq (E_{\rm P2} \leq E_{\rm C2}$ 
{\rm or} $E_{\rm C2} < E_{\rm P2}$), we choose P1 and return it to 
S$^{\prime}_k$ together with a new individual. 
\end{enumerate}

In case (a), the number $n_k$ of the individuals of S$^{\prime}_k$ increases, 
whereas in case (b) it decreases. We add a new individual to S$^{\prime}_k$ 
when the number becomes less than two. 

\item When the best individual ($E_{\rm C1} < \min\{E_{\rm P}\}$) is created, 
we add it to a different local population, which is randomly selected. 

\item We stop searching for the ground state if the best individual is not created during 100 generations 
($\Delta t = 100$). 
Otherwise, return to (1).
\end{enumerate}

Figures 1 and 2 show examples of the ground-state search process in a typical 
sample. In Fig. 1, the lowest energies $E_n(t)$ for eight different 
populations (eight trials) are presented as functions of the generation $t$. 
We see that, for each population, stepwise decreases follow an abrupt decrease 
at $t \sim 0$; its interval $\Delta t$ increases with $t$. 
Finally, $E_n(t)$ reaches $E_G$ at $t \sim 100$ 
(further $\Delta t = 100$ generation is necessary to confirm that 
no more stable state will appear in the population). 
In Fig. 2, the distance $\Delta S_{n,g}(t)$ between the spin 
configuration \{$\bm{S}_i^{(n)}$\} with $E_n(t)$ and the ground-state spin 
configuration is presented in the same search process. 
Each has a value of $\Delta S_{n,g}(t) \sim 0.6\;(\sim 0.7\times \Delta 
S_{\infty})$ at $t \sim 0$ and changes irregularly with increasing $t$ 
around this value of 0.6 until the system becomes near the ground state. 
These suggest that the system reaches the ground state via local 
minimum states, the spin configurations of which are considerably different 
from the ground-state spin configuration. 
That is, the ground-state spin configuration will suddenly appear in this 
search process.

\begin{figure}[t]
\begin{center}
\includegraphics[width=8cm]{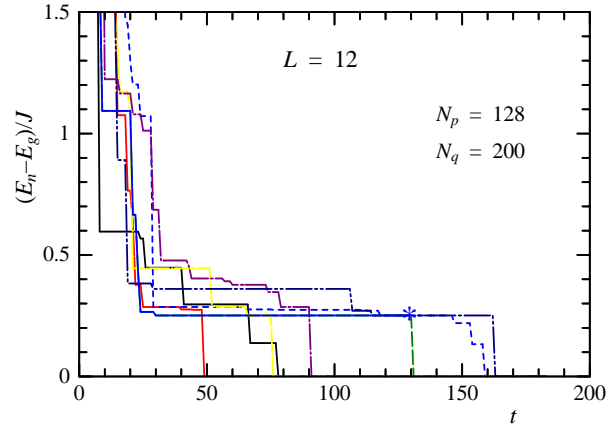}
\end{center}
\caption{Lowest energy $E_n$ for the $n$th trial in a typical sample 
of the $\pm J$ Heisenberg model as a function of the generation $t$, where 
$E_G$ is a tentative ground-state energy obtained beforehand, 
applying the same method with a larger number of the local populations 
of $N_p = 256$. 
Note that results for eight trials are shown, one of which (described by *) 
fails to get the ground-state energy because $\Delta t > 100$.}
\label{f1}
\end{figure}

\begin{figure}[t]
\begin{center}
\includegraphics[width=8cm]{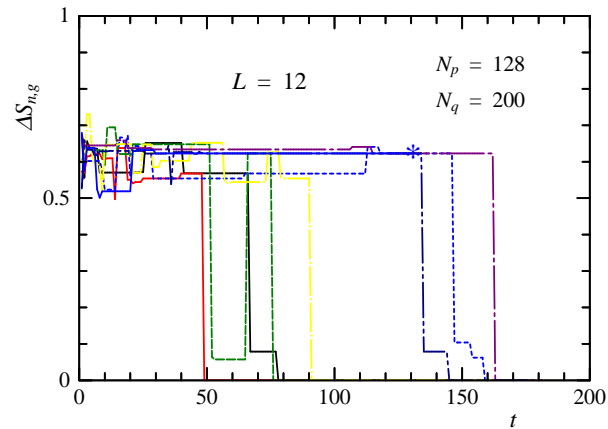}
\end{center}
\caption{Distance in the spin configuration $\Delta S_{n,g}$ 
as a function of the generation $t$ in the same ground-state search 
process presented in Fig. 1. }
\label{f2}
\end{figure}

The PfGA requires only two parameters: the number of the local populations 
$N_p$ and the number of quenching steps $N_q$. In contrast to the GAI, 
the generation number $\nu$ results from the evolution of the population. 
It was found that, when $N_p$ and $N_q$ are fixed, the average number 
$\bar{\nu}$ slowly increases with the linear size $L$, e.g., when $N_p=128$ 
and $N_q = 100$, $\bar{\nu}$ = 154(16), 149(20), 200(20), 219(26) and 287(30), 
respectively, for $L$ = 8, 10, 11, 12 and 13. 
It was also found that, for a fixed size $L$, $\bar{\nu}$ decreases slowly as 
$N_p$ and/or $N_q$ increases. 
Although the total number of the individuals in the population S 
changes as the population evolves and often becomes considerably greater 
than $2N_p$, the crossover time is $2N_p$ at every generation. 
Therefore, the computer CPU time $t_{\rm II}$ is estimated as 
\begin{equation}
  t_{\rm II} = A_{\rm II} \times 2N_p \times \bar{\nu} \times N_q 
                           \times({\rm lattice\ size}\ N), 
\end{equation}
with $\bar{\nu} = 150 \sim 300$ for $L \leq 14$. The constant $A_{\rm II}$ 
is independent of other parameters and $A_{\rm II} \sim A_{\rm I}$. 
We have performed the same test in the $\pm J$ Heisenberg model; its results 
are presented in Table II. 
It is readily apparent that $t_{\rm II}/t_{\rm I} = 1/4 \sim 1/2$ for 
obtaining the ground state with the same validity; using GAII, we can 
extend the treatable lattice size using a personal computer from $L = 12$ 
to $L = 13$. 
We have calculated the ground state energy per spin, $E_G(L)$, for the 
$L\times L\times L$ ($L \leq 13$) lattice and present them in Table III. 
Here, the parameter set of ($N_p, N_q$) has been chosen such that the 
search ratio becomes greater than 0.90. The numbers of samples are about 
4000 for smaller lattices and about 500 for larger lattices. 
We have estimated the ground state energy of the model, 
$E_G (\equiv E_G(\infty))$, by using data for  $8 \leq L \leq 13$ with 
an extrapolation function $E_G(L) = E_G + a/L^{\lambda}$, 
and add it in the same table. Note that the value of $E_G/J=-2.0432\pm0.0015$ 
is a little lower than that of $E_G/J \sim -2.0411$ 
estimated by using data for smaller lattices of 
$4 \leq L \leq 11$\cite{Takahashi}. 
Further studies are necessary to settle the value of $E_G$.

We have developed a new genetic algorithm (GAII) by combining the triadic 
crossover of the GAI and the parameter free genetic algorithm. 
The GAII further improves the ground-state searching efficiency. 
It reduces the number of parameters from four ($M_i, \nu_0, N_q$ and $\mu$ 
in the GAI) to two ($N_p$ and $N_q$). This is also the merit of the method 
because the determination of those parameters is an important 
but tedious task for application to different problems.

\begin{table*}
\caption{\label{tabone} Ratios and their standard variances of success 
in searching for the ground-state energy of the $\pm J$ model on 
$L \times L \times L$ lattice using the GAII, 
where $N_p$ and $N_q$ respectively denote the local population and 
number of quenching steps. These are estimated for 10 runs for 
16 samples. }
\vspace{0.3cm}
\begin{center}
\begin{tabular}{@{\hspace{\tabcolsep}\extracolsep{\fill}}r|ccccccc} \hline
$L(N_q) \setminus  N_p$  & 16 & 32 & 64 & 128 & 256 & 512  & 1024   \\  \hline
8  (100) &0.93(.13)&1.00(.00)&---      &---      &---&---&---       \\
10 (100) &---      &0.80(.18)&0.90(.14)&0.98(.03)&---&---&---       \\
11 (100) &---      &---&0.59(.20)&0.85(.17)&0.92(.13)&---&---       \\
   (200) &---      &---&0.83(.14)&0.80(.17)&0.99(.03)&---&---       \\
12 (200) &---      &---&---&0.85(.17)&0.92(.12)&1.00(.00)&---       \\
13 (400) &---      &---&---&---      &0.74(.22)&0.86(.20)&0.93(.13) \\
14 (400) &---      &---&---&---      &---      &0.71(.19)&0.81(.23) \\
\hline
\end{tabular}
\end{center}
\end{table*}

\begin{table*}
\caption{\label{tabone} The ground state energy per spin $E_G(L)$ of 
the $\pm J$ model on the $L\times L\times L$ lattice obtained by GAII. 
$E_G (\equiv E_G(\infty))$ is estimated by using data for $8 \leq L \leq 13$.} 
\vspace{0.3cm}
\begin{center}
\begin{tabular}{@{\hspace{\tabcolsep}\extracolsep{\fill}}r|ccccccc} \hline
$ L$  & 8 & 10 & 11 & 12 & 13 & $\infty$ \\  \hline
$E_G(L)/J$  &-2.0341(2)&-2.0376(2)&-2.0392(3)&-2.0393(3)&
-2.0403(3)& -2.0432(15) \\
\hline
\end{tabular}
\end{center}
\end{table*}

\section{Stiffness of the Ground State}

The most interesting ground-state property of the system is the stiffness of 
the model at $T = 0$. 
The stiffness constant $\theta$ is estimated from the domain-wall (DW) 
energy $\Delta E_L$ of the $L \times L \times L$ lattice; 
\begin{equation}
\Delta E_L \propto JL^{\theta},
\end{equation}
for $L \rightarrow \infty$. 
When $\theta > 0$, the SG phase transition occurs at a finite temperature, 
although no phase transition occurs when $\theta < 0$. 
A mysterious problem exists, by which $\theta$ depends on the estimation 
method of $\Delta E_L$. Two methods are typically used: 
One is the periodic-antiperiodic (P-AP) method, in which
$\theta$ is estimated as 
$\theta = -(0.65\sim1.0)$\cite{Banavar,McMillan1,Kawamura2}. However, 
this value was shown to be lattice size range dependent.\cite{Matsubara1}. 
The other is the open-boundary-twist (OB-Twist) 
method \cite{MSS,Endoh1,Endoh2}, by which 
one obtains $\theta = 0.5\sim1.0$ depending on the 
twisting manner.\cite{Matsubara1} 
That is, the former method predicts the absence of the phase transition at 
a finite temperature; the latter predicts its presence. 
Unfortunately, these estimations are given in small lattices of $L \leq 8$. 
The question is whether or not the two methods engender the same result for 
$L \rightarrow \infty$. Here, we reexamine $\theta$ in larger lattices 
using the GAII.

\subsection{P-AP method}
Using this method, one usually considers a simple cubic lattice of 
$L \times L \times L$. Here, we consider lattices with different aspect 
ratios $L \times L \times rL$ ($r \geq 1$) because the value of $\theta$ 
of the system was suggested to be properly estimated in lattices with 
a large ratio $r$\cite{Aspect,Aspect2}. 
We consider the DW energy $\Delta E_{r,L}$ which is defined as the difference 
in the ground-state energy between the two lattices {\it A} and {\it B} 
with the same bond distribution but with different boundary conditions. 
That is, for lattice {\it A}, a periodic boundary condition is applied for 
every direction; for lattice {\it B}, an antiperiodic boundary condition 
is applied for the $z$-direction and a periodic boundary condition 
for the $x$- and $y$-directions. 
The GAII is applied to both lattices {\it A} and {\it B} to estimate the 
respective ground-state energies $E_{r,L}^{(A)}$ and $E_{r,L}^{(B)}$. 
The DW energy is calculated as [$\Delta E_{r,L}=|E_{r,L}^{(A)} 
- E_{r,L}^{(B)}|$], where [$\cdots$] denotes the sample average. 
The parameter set of ($N_p, N_q$) and the numbers of the samples are the 
same as described in Sec. 2.2. 
Figure 3 shows [$\Delta E_{r,L}$] for different $r$ as functions of $L$ in 
{\it a log-log form}, the slope of which gives the stiffness constant 
$\theta$. 
We see that [$\Delta E_{r,L}$] for different $r$ shows different $L$ 
dependences; for $r = 1$, it decreases with increasing $L$, but its 
decrement becomes smaller and seems to converge to a finite, nonzero value 
for $L \rightarrow \infty$; for $r = 2$, it is almost independent of $L$; 
for $r \geq 3$, it increases with $L$ and seems to converge to a finite value. 
That is, the results imply that $\theta=0$, irrespective of the aspect 
ratio $r$. 
Then using an extrapolation function 
$[\Delta E_{r,L}]=[\Delta E_{r,\infty}] +a/L+b/L^2$ for $L\geq 4$, 
we estimate the convergence value for each of $r$: 
$[\Delta E_{r,\infty}]/J = 0.95\pm0.06, 0.49\pm0.06, 0.36\pm0.06$ and 
$0.22\pm0.08$ for $r = 1, 2, 3$ and 4, respectively. 
As expected, results fit well like in a ferromagnetic model: 
$[\Delta E_{r,\infty}] \propto [\Delta E_{1,\infty}]/r$. 
Therefore, we suggest that $\theta = 0$ in the P-AP method, 
contrary to previous estimations of $\theta < 0$. 

\begin{figure}[t]
\begin{center}
\includegraphics[width=8cm]{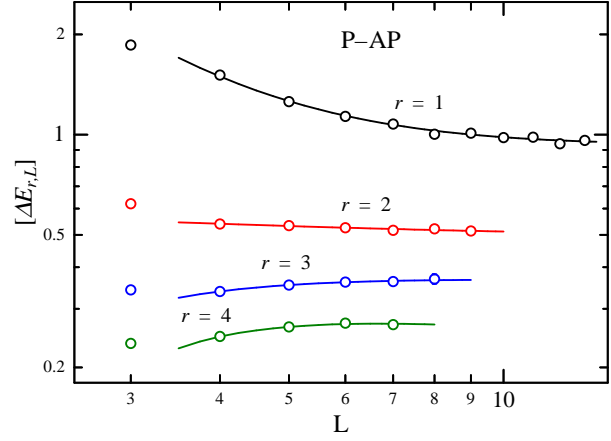}
\end{center}
\caption{DW energy $\Delta E_{r,L}$ of the $L \times L \times rL$ 
lattice using the P-AP method. Lines are fitted ones using the function 
described in the text. 
 }
\label{f3}
\end{figure}

\subsection{OB-twist method}
We consider the $L \times L \times (rL+1)$ lattice with periodic boundary 
conditions for the $x$- and $y$-directions and the open boundary condition 
for the $z$-direction (the lattice has two opposite surfaces, 
$\Omega_1$ and $\Omega_{rL+1}$). We consider a twist energy, 
which is calculated as follows. 
We first determine the ground state with an energy $E_{r,L}$; then, under 
the condition that all the spins on $\Omega_1$ are fixed, all the spins on 
$\Omega_{rL+1}$ are twisted (rotated) at $\phi$ around the $z$-axis and 
the lowest energy $E_{r,L}(\phi)$ is calculated; the twist energy is given 
as the difference between these two energies: 
$\Delta E_{r,L}(\phi) = E_{r,L}(\phi)-E_{r,L}$. 
Using the GAII with similar conditions to those used in the P-AP method, 
we calculate $[\Delta E_{r,L}(\phi=\pi/2)]$ for $r = 1, 2, 4$. The results are 
presented in Fig. 4 in {\it a log-log form}. In contrast to the results in 
the P-AP method, data for each $r$ seem to lie on a straight line, suggesting 
that $[\Delta E_{r,L}(\pi/2)] \propto L^{\theta}$. 
Their lines' slopes are almost identical, 
which indicates that $\theta$ is independent of the aspect ratio $r$. 
Then we fit them using a stiffness constant of $\theta = 0.62$, which was 
estimated previously in smaller lattices ($L \leq 8$) with 
$r = 1$.\cite{Matsubara1} 
The quality of the results is very good, as depicted in Fig. 3. 
Consequently, we expect that $\theta \sim 0.62$ in the OB-twist method.

\begin{figure}[t]
\begin{center}
\includegraphics[width=8cm]{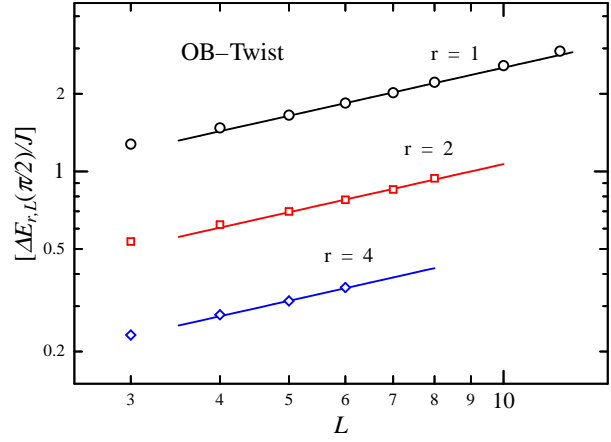}
\end{center}
\caption{Twist energies $\Delta E_{r,L}(\pi/2)$ of the 
$L \times L \times rL$ lattice with open boundaries as functions of $L$. 
Lines are fitted ones using the stiffness exponent $\theta=0.62$, which was 
given using smaller lattices ($L\leq8$) with $r = 1$.\cite{Matsubara1}
} 
\label{f4}
\end{figure}

\subsection{Remarks related to the stiffness constant $\theta$}

We estimated the stiffness constant $\theta$ in larger lattices 
using the P-AP and OB-twist methods and obtained different values, 
respectively, of $\theta=0$ and $\theta \sim 0.62$. 
That is, $\theta$ of this model depends on the estimation method, 
in contrast to a ferromagnetic Heisenberg model for which $\theta$ is 
a universal constant of $\theta = d-2$ in the $d$-dimensional system. 
We consider $\theta$ in those methods. 

In the OB-twist method, the meaning of $\Delta E_{r,L}(\phi)$ is clear. 
We consider the stiffness of the ground state of the open system itself; 
the depth or steepness of the ground-state valley in the energy landscape 
of the system. 
In this case, the twist energy $\Delta E_{r,L}(\phi)$ is surely a lifted 
energy brought by a perturbation around the ground state. 
That is, the result for $\theta \sim 0.62$ suggests that, once the system 
falls into a ground-state valley at very low temperatures, 
it slightly escapes from the valley.

On the other hand, in the P-AP method, the calculation of $\Delta E_{r,L}$ 
comes from an application of the renormalization-group idea; one evaluates 
the effective coupling $\tilde{J}_L (\sim JL^{\theta})$ between block spins 
of the linear dimension $L$ generated by renormalization. 
That is, one assumes that $\tilde{J}_L \sim \Delta E_{r,L}$. 
However, the meaning of $\Delta E_{r,L}$ calculated in this method has 
remained unclear.\cite{Endoh1,Endoh2,KA} 
Our findings of $\theta=0$ and $[\Delta E_{r,\infty}]
\propto[\Delta E_{1,\infty}]/r$ seem to reveal its meaning. 
We consider $\Delta E_{r,L}$ on the basis of the DW argument in the 
ferromagnetic Heisenberg model. 
The difference $\Delta E_{r,L}$ arises from the difference in the boundary 
condition between lattices {\it A} and {\it B}. That is, the 
difference in the energy between two lattices with different boundary 
bonds $\{J_{(i,1),(i,rL)}\}$ and $\{-J_{(i,1),(i,rL)}\}$. 
We first consider the open lattice with $\{J_{(i,1),(i,rL)} = 0\}$ for which 
the ground-state spin configuration and its energy are described, 
respectively, as $\{\bm{S}^0_i\}$ and $E^0_{r,L}$. 
Coming back to lattice {\it A}, some boundary bonds will favor 
$\{\bm{S}^0_i\}$; others will obstruct it, giving the resulting energy 
$E^{(A)}_{r,L}$.
We respectively denote the former bonds as right ($r$) -bonds and the latter 
bonds as wrong($w$)-bonds and their numbers as $n_r$ and $n_w$. 
As in the ferromagnetic case, this operation will change the energy 
of the lattice as $O(J/rL)$ per chain along the $z$-direction for larger $L$, 
$E^{(A)}_{r,L} - E^0_{r,L} \propto - (J/rL) \times (n_r-n_w)$. 
In lattice {\it B}, the roles of the $r$- and $w$-bonds are 
reversed and $E^{(B)}_{r,L} - E^0_{r,L} \propto - (J/rL) \times (n_w-n_r)$. 
Then we expect $\Delta E_{r,L} \propto JL^{\frac{d-3}{2}}/r$ for $L \gg 1$, 
because $|n_r-n_w|\sim L^{\frac{d-1}{2}}$ in the $d$-dimensional lattice. 
That is, $\theta=0$ in $d=3$ and $[\Delta E_{r,\infty}]
\propto[\Delta E_{1,\infty}]/r$, as was obtained in this study. 
To examine the speculation, we make an additional calculation putting all the 
surface bonds as $r$ bonds, as in the usual DW argument in the ferromagnetic 
case: $\{J_{(i,1),(i,rL)} = Sign(\bm{S}^0_{i,1}\bm{S}^0_{i,rL})J\}$. 
We call this boundary condition an optimized boundary condition and the 
boundary condition with $\{-J_{(i,1),(i,rL)}\}$ an antioptimized 
boundary condition, and the calculation of $\Delta E_{r,L}$ 
in these boundary conditions as an optimized-antioptimized (Opt-AOpt) method. 
Results of $\Delta E_{r,L}$ for $r = 1,$ and 2 are shown in Fig. 5. 
In fact, $\Delta E_{r,L}$ increases with $L$ for larger $L$. 
Interestingly, their slopes become larger with increasing $L$, implying 
that they reach 1 for $L \rightarrow \infty$, as suggested by the argument. 
Further calculations are necessary to estimate the value of $\theta$ for 
each $r$ and thereby its $r$-independent value. 
Nonetheless the Opt-AOpt method implies that $\theta > 0$ in constrast with 
$\theta = 0$ in the P-AP method. 
We suggest, therefore, that the $\theta$ in the P-AP method is not 
a universal constant that describes the effective coupling $\tilde{J}_L$, 
but a characteristic value that describes the boundary condition. 
It turns out the results imply that a strong correlation exists between 
the spin configurations on two opposite surfaces of the open lattice.

Hence, we suggest that the result of $\theta = 0$ in the P-AP method and 
that of $\theta \sim 0.62$ in the OB-twist method indicate the same thing that 
the ground state of the open system is stable against a weak perturbation. 

\begin{figure}[t]
\begin{center}
\includegraphics[width=8cm]{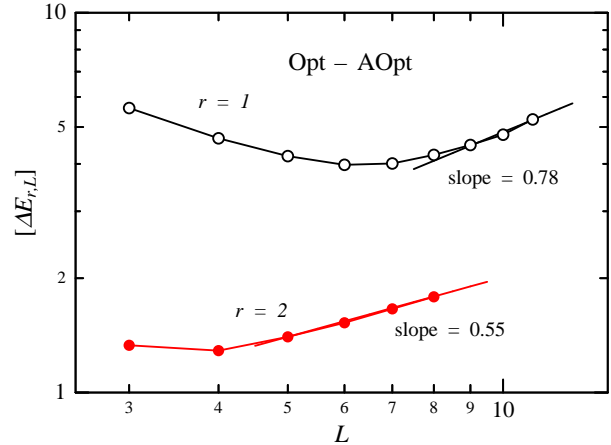}
\end{center}
\caption{The DW energy $\Delta E_{r,L}$ of the $L \times L \times rL$ 
lattice using the Opt-AOpt method. 
 }
\label{f5}
\end{figure}

\section{Conclusions}
We developed a genetic algorithm (GA) in the Heisenberg model. 
We first show that the triadic crossover in the Ising model 
also works well in the Heisenberg model. 
Then, combining it with a parameter-free genetic algorithm, 
we proposed a useful algorithm for the Heisenberg spin-glass model. 

Using the algorithm, we examined the ground-state stiffness 
of the $\pm J$ Heisenberg model in three dimensions. 
Results showed the stiffness constant of $\theta = 0$ in the 
periodic-antiperiodic method, and that of $\theta \sim 0.62$ in the 
open-boundary-twist method. 
The origin of the difference in $\theta$ between these methods was discussed. 
We suggested that both results show the same thing: the open system's ground 
state is stable against a weak perturbation. An interesting issue whether 
or not the values of the chirality stiffness constant and the spin glass 
stiffness constant are the same will be discussed in a separate paper.

\section*{Acknowledgements}
 The authors would like to thank Professor T. Shirakura, 
Professor K. Sasaki, and Dr. M. Sasaki for their useful discussions.




\begin{thebibliography}{99}
\bibitem{Comm_Traidic} We can change this criterion of $r_i^{(P1)} > 0$ or 
$r_i^{(P1)} < 0$ to $r_i^{(P1)} > r_i^{(P2)}$ or $r_i^{(P1)} < r_i^{(P2)}$, 
but the efficiency hardly improves.
\end{thebibliography}
\end{document}